\documentclass[aps,10pt,twocolumn,preprintnumbers,amsmath,amssymb]{revtex4-1}
\usepackage{graphicx}
\usepackage{epstopdf}
\usepackage[utf8x]{inputenc}
\usepackage{amsmath}
\usepackage{url}
\usepackage{hyperref}
\usepackage{color}
\usepackage{siunitx}
\def\UOM#1{\,\hbox{#1}}
\definecolor{orange}{rgb}{1,0.5,0}
\definecolor{grey}{rgb}{0.4,0.4,0.4}
\definecolor{green}{rgb}{0.13,0.54,0.13}

\begin{document}

\title{Accurate phase diagram of tetravalent DNA nanostars}

\author{Lorenzo Rovigatti} 
\author{Francesca Bomboi} 
\affiliation{{Dipartimento di Fisica, {\em Sapienza} Universit\`a di Roma, Piazzale A. Moro 2, 00185 Roma, Italy}}
\author{Francesco Sciortino} 
\affiliation{{Dipartimento di Fisica, {\em Sapienza} Universit\`a di Roma, Piazzale A. Moro 2, 00185 Roma, Italy}}

\begin{abstract}
We evaluate, by means of molecular dynamics simulations employing a realistic DNA coarse-grained model, the phase behaviour and the structural and dynamic properties of tetravalent DNA nanostars, i.e.  nanoconstructs 
completely made of DNA. We find that, as the system is cooled down, tetramers undergo a gas--liquid phase separation in a region of concentrations which, if the difference in salt concentration is taken into account, is comparable with the  recently measured experimental phase diagram [S. Biffi \textit{et al}, Proc. Natl. Acad. Sci, \textbf{110}, 15633 (2013)].  We also present a mean-field free energy for modelling the phase diagram  based on the 
bonding contribution derived by Wertheim in his studies of associating liquids. Combined with 
mass-action law expressions appropriate for DNA binding and a numerically evaluated reference free energy,
the resulting free energy qualitatively reproduces the numerical data.
Finally, we report information on the nanostar structure, e.g. geometry and flexibility  of the single tetramer 
and of the collective behaviour, providing a useful reference for future small angle 
scattering experiments, for all investigated temperatures and concentrations.
\end{abstract}

\maketitle

\section{Introduction}
In the colloidal realm it is possible, in principle, to fabricate building blocks with desired properties. As a consequence, 
by carefully designing the inter-particle interactions,  materials with novel and technologically relevant properties
could be assembled.  In the last decade a lot of effort has been put into the development of new techniques aimed at the synthesis of these new generation of colloidal particles. These recent advances allow for the creation of an incredible variety of anisotropically interacting building blocks~\cite{Glotz_Solomon_natmat}. The anisotropy can arise from shape, surface patterning, form of the interactions or a combination thereof. Examples are colloidal cubes~\cite{sun_nanocubes,philipse_cubes}, Janus particles~\cite{Janus,janusweitz}, triblock Janus particles~\cite{granick_janus}, patchy particles~\cite{Manoh_03,Cho_05,dayang_patchy}, magnetic spheres~\cite{klokkenburg_dipolar_spheres} and many others. Despite the progress, fabricating monodisperse, anisotropic nano-- or micro--sized particles with tunable interactions in bulk quantities such that the 
collective behavior can be explored is still limited to very few cases. A fortunate case is offered by DNA 
constructs, supramolecules of namometric size, which exploit 
 the high selectivity of the Watson-Crick base pairing mechanism. 
Recent developments in DNA synthesis and nanotechnology have made it possible to exploit DNA as a building block to produce 2D and 3D crystals~\cite{winfree_tiles,seeman_dna_review_2003,zheng2009molecular,seeman_dna_review_2010}, hierarchical self-assembly of tiles~\cite{winfree_tiles} and self-assembly of strands into large, albeit inherently finite, structures, i.e. DNA origamis~\cite{rothemund_origami}. DNA can also be used to functionalise colloids by grafting single-stranded DNA (ssDNA) molecules on the surface of the particles~\cite{mirkin_96,dnastarr,geerts_dna}. By choosing strands terminating with complementary sequences, the particles can bind to each other via DNA hybridisation and self-assemble into disordered or ordered structures~\cite{dnastarr,langmuir,park2008dna,nykypanchuk2008dna,frenkel_dna_colloids,eiser_review}.

\begin{figure}
\centerline{\includegraphics[width=1\linewidth]{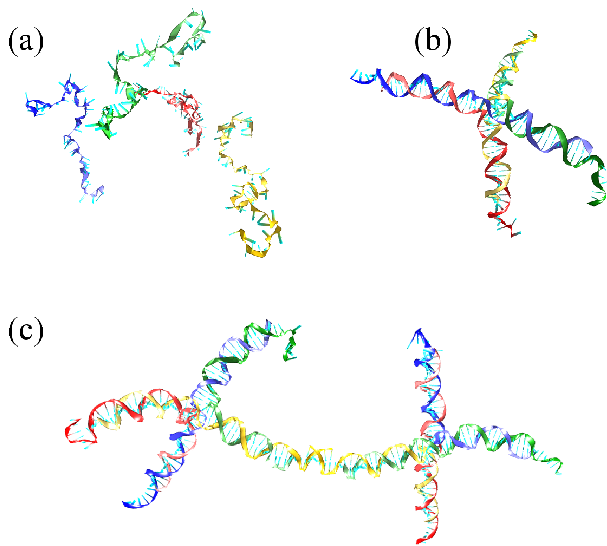}}
\caption{The investigated two-step self-assembly process. (a) At very high temperatures ($T \gtrsim 80\; ^\circ C$) there are no hydrogen bonds between the strands. (b) For intermediate $T$ ($50\; ^\circ C \lesssim T \lesssim 80\; ^\circ C$) single strands start to hybridise and tetramers are formed. (c) Upon further cooling, tetramers are linked together by hybridisation of the sticky ends. Each different strand color corresponds to a different sequence (see Table~\ref{tbl:tetra}). All the sticky ends have the same palindromic sequence and hence they can bind to each other regardless of the strand they are part of.}
\label{fig:tetra_intro}
\end{figure}

Very recently, the phase diagram of a system of limited-valence particles, consisting entirely of
 DNA strands,  has been experimentally measured.   Particles are composed
by different single strands, properly designed to self-assemble into nanostars at a desired temperature via the Watson-Crick pairing.  A  short single-strand sequence is left at the end of each arm 
to provide bonding betwen different constructs.  Figure~\ref{fig:tetra_intro} shows a sketch of the two-step self-assembling process that the strands undergo. At high temperature $T$, above the melting temperature of single tetramers, all the DNA is in single-stranded form and there are no DNA constructs. As the system is cooled down, tetramers start to form, but they weakly interact with each other. When $T$ is significantly lower than the melting temperature of single tetramers, DNA constructs start to form bonds through the tips of the arms. These \textit{sticky ends} are self-complementary and allow for inter-tetramer bonding. These star-shaped constructs
with  three (DNA trimers) or four (DNA tetramers) arms have been studied to investigate the effect of lowering the valence on the location of the gas--liquid coexistence region of the phase diagram~\cite{biffi_dna}.

The experiments have confirmed previous theoretical and numerical predictions, in that 
the gas-liquid instability region has been shown to shrink and move to low temperatures and concentrations as the valence decreases~\cite{biffi_dna}, giving rise to very low-density, open equilibrium networks, the so-called empty liquids~\cite{bian,zaccajpcm,barbaranatmat}.  These DNA constructs can be considered as a realisation of patchy particles, which, in the last years, have been used as simple model systems to predict the existence and investigate the properties of exotic states of matter such as empty liquids~\cite{bian}, open crystals~\cite{doye_tetra,flavio_nature_comm}, reentrant gels~\cite{sandalo_gel} and self-assembling systems~\cite{reinhardt,closed_loop_prl}.

The phase behavior of a very primitive model for  DNA tetramers, composed by
four DNA strands attached to a central core,  has been previously numerically investigated in a series of studies~\cite{dnastarr,langmuir,starr_pnas}. This simple model predicted the existence of a gas-liquid unstable region at low densities, followed at larger densities by a gel phase in which tetramers form an extended network of fully bonded particles. In this article we  improve on these previous studies by  employing a more realistic coarse-grained DNA model recently developed by Ouldridge \textit{et al.}~\cite{ouldridge_prl,ouldridge_jcp,tesi_tom}. This model
is quite accurate in predicting the melting temperature of DNA sequences, the correct 
single and double strand persistence lengths in addition to  base pair selectivity. It thus provides 
an excellent tool for DNA nanotechnology applications.  We  simulate the same DNA sequences
used in the experimental study of Ref.~\cite{biffi_dna},  offering a microscopic description of
the collective phenomena experimentally observed. More specifically, we make use of extensive molecular dynamics simulations on graphic processor units (GPUs) to simulate bulk systems of DNA tetramers at different concentrations and temperatures. We focus on the region of the phase diagram where the DNA constructs undergo a gas-liquid-like phase separation and we study the structure and the dynamics as the system approaches the spinodal curve.
As shown in the following sections, the use of a realistic model of DNA, consistent with the experimental melting curves as parametrized in the SantaLucia framework~\cite{santalucia}, offers the possibility of developing a theoretical formalism for evaluating the gas-liquid phase coexistence which can be first tested against the numerical findings and then against  the experimental results. We do so  by combining the
bonding free energy proposed by Wertheim in his studies of associating liquids~\cite{Werth1,Werth2,Werth4}, the
Santa Lucia mass-action law expressions and a numerically-evaluated reference free energy at the virial
level. The resulting expression properly models the numerical data and the experimental results. 

\section{Numerical Methods}

\begin{table}
\centering
\begin{tabular}{|c|}
\hline
\textbf{Strand sequences} \\
\hline 
\texttt{\color{red}CTACTATGGCGGGTGATAAA\color{grey}AA\color{red}CGGGAAGAGCATGCCCATCC\color{grey}A\color{black}CGATCG}\\
\texttt{\color{blue}GGATGGGCATGCTCTTCCCG\color{grey}AA\color{blue}CTCAACTGCCTGGTGATACG\color{grey}A\color{black}CGATCG}\\
\texttt{\color{green}CGTATCACCAGGCAGTTGAG\color{grey}AA\color{green}CATGCGAGGGTCCAATACCG\color{grey}A\color{black}CGATCG}\\
\texttt{\color{orange}CGGTATTGGACCCTCGCATG\color{grey}AA\color{orange}TTTATCACCCGCCATAGTAG\color{grey}A\color{black}CGATCG}\\
\hline
\end{tabular}
\caption{The strand sequences designed to self-assemble into tetramers. We use the same sequences used in the experimental work of Biffi \textit{et al.}~\cite{biffi_dna}. Coloured nucleotides form the double-stranded parts of the arms, spacers are in grey and sticky ends are in black.}
\label{tbl:tetra}
\end{table}

The interaction forms and parameters of the coarse-grained DNA model we employ, oxDNA, are chosen to
reproduce structural and thermodynamic properties of both single- and double- (dsDNA) stranded 
molecules of DNA in B-form. All interactions between nucleotides are pairwise, continuous and differentiable. The only sequence-dependence 
of the model we use here is in the specificity of the Watson-Crick bonding. A new version of the 
model, which features a limited sequence dependence also in the interaction strengths, has been 
recently developed~\cite{dna_code}.

The interactions between nucleotides account for excluded volume, backbone connectivity, Watson-Crick 
hydrogen bonding, stacking, cross-stacking and coaxial-stacking. The interaction parameters have been adjusted in order to be consistent with experimental data~\cite{ouldridge_jcp,santalucia,Holbrook}.
In addition, the model is parametrised for a specific value of salt molarity ($0.5 \UOM{M} $ NaCl). In computing the concentration $c$ we use an average nucleotide mass of $m = 330 \; Da$. A code implementing the oxDNA model is freely available on the web~\cite{dnawebsite}.

We perform brownian simulations and investigate systems consisting of $100$ DNA tetramers. Each tetramer is formed by four strands, each composed of $49$ nucleotides. Strand sequences are presented in Table~\ref{tbl:tetra}. Each sequence can be divided into three regions, separated by one or two
nucleotides that act as spacers and provide flexibility to the centre of the tetramer and to the sticky ends.
The first two regions are $20$-base long and are designed to form the double-stranded parts of the arms. The third region, composed of $6$ nucleotides, is
identical in all four sequences and self-complementary. This final sequence
functions as a sticky end, allowing for inter-tetramer bonds. The difference in
length between the double-stranded arms and the sticky ends provides a
separation in between the temperature at which tetramers assemble and the
temperature at which they start to form a network.

The system contains a grand total of $19600$ individual rigid bodies, interacting through a very complicated and numerically-intensive potential. By harvesting the computational power of modern GPUs we are able to boost up performances by a factor $40-50$, if compared with CPU simulations. Equilibration and production simulations have been run up to $10^9$ MD steps for each state point, corresponding to $\sim 10 \; \si{\micro\second}$ of real time. Without the speed-up provided by the GPU code, each investigated system would have taken up to decades on a single CPU core.

We construct an initial tetramer by putting a high concentration ($c \approx 100\; \si{mg/ml}$) of an equal number of constituent strands in the simulation box. We then simulate the system at $T = 60 \; ^\circ C$, below the single tetramer melting temperature but above the temperature at which tetramers start to assemble. As soon as a defect-free, complete tetramer is formed we extract and replicate it to generate initial configurations at different concentrations.

\begin{figure}
\centering
\includegraphics[width=1\linewidth]{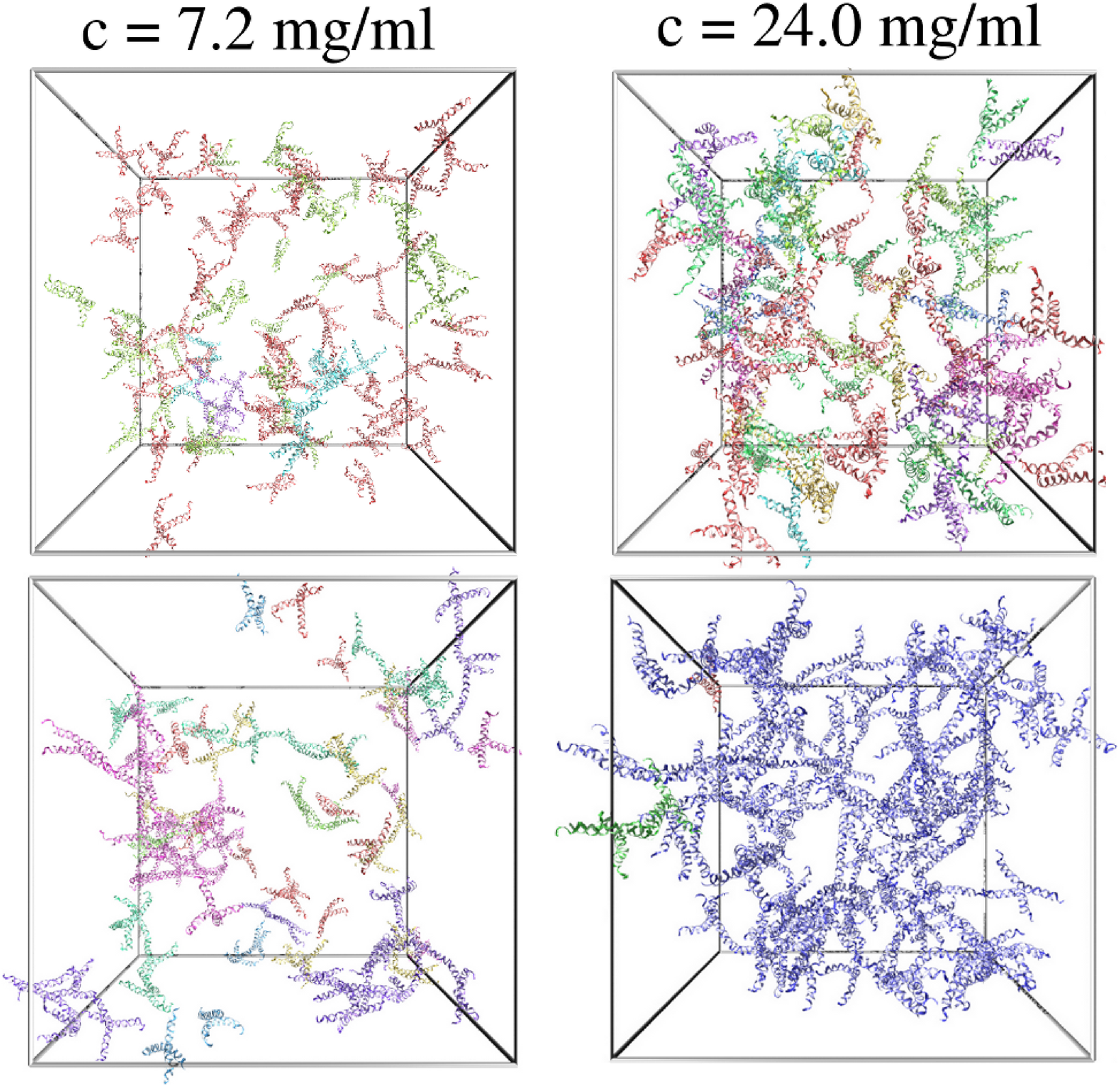}
\caption{Snapshots taken from simulations at concentration $c = 7.2 \;\si{mg/ml}$ (left panels) and $c = 24.0 \; \si{mg/ml}$ (right panels) at high ($T = 48^\circ \;C$) and low ($T = 39^\circ \;C$) temperature. Tetramers are coloured according to the size of the cluster they are part of. At high temperatures (top panels) there are mostly small clusters (depicted in red, green and magenta). Upon lowering $T$, the system starts to form clusters. At the lowest $T$ (bottom panels) the system at low concentration is inhomogeneous whereas, at high concentration, a percolating cluster spans the whole simulation box.}
\label{fig:tetra_snaps}
\end{figure}

We simulate systems in a temperature range $39\; ^\circ C \leq T \leq 48\; ^\circ C$  for five different concentrations, namely $7.2 \; \si{mg/ml}$, $12.2 \; \si{mg/ml}$, $16.3 \; \si{mg/ml}$, $20.0 \; \si{mg/ml}$ and $24.0 \; \si{mg/ml}$, which correspond to box sizes of, respectively, $L \approx 114$, $95.5$, $86.4$, $81.1$ and $76.3 \; \si{\nano\metre}$. Fig.~\ref{fig:tetra_snaps} shows images of representative configurations   at different state points.

In the following analysis, we consider two tetramers as bound to each other if their sticky ends share at least three bonded nucleotides. Two nucleotides are regarded as bonded if the interaction energy term due to hydrogen bonding is less than $-0.1 \epsilon$, where $\epsilon$ is the energy unit of the model. We note that changing either threshold does not significantly affect the results, since two sticky ends share nearly always either zero or all six fully-formed base-pairs. This is to be expected, since free-energy profiles of duplex formation below the melting temperature show that fully-hybridised structures are by far the most stable configurations if, as in the present case because of dangling-end stabilisation, the fraying effect is not relevant~\cite{ouldridge_jcp,depablo_biophysics}. 

We note that one way of producing empty liquids~\cite{bian,zaccajpcm,barbaranatmat}, is to employ valence-limited particles that fulfil the single-bond-per-patch condition. For the system studied in this work the very nature of DNA hybridisation satisfies this condition, although the large flexibility of the arms may enable multiple bonding between neighbours. In order to check if this is the case we compute the fraction of multiple bonds, i.e. the number of bonds that connect the same pair of tetramers over the total number of bonds. The results show that, at the lowest studied temperature, there are no more than a few percent (up to $5\%$) of multiple bonds. Such a value should not significantly affect the network topology of the resulting gel.

In the following, if not otherwise stated, we use the centres of mass of tetramers to carry out analyses of the configurations.

\section{Theory}

In order to theoretically estimate the location of the gas--liquid phase transition we combine Wertheim's theory~\cite{Werth1,Werth2,Werth4} with  accurate  mass action law describing DNA binding. The basic assumptions of Wertheim's theory are that (i) all  bonds are independent, i.e. the state of a patch, being it either bonded or unbonded, does not depend on the state of any other patch, and (ii) no bond loops form  in the finite size aggregates (i.e. no intra-cluster bonds). Under these hypothesis, the formation of one bond in the system requires the decrease by one of the number of clusters. Indeed, the Helmholtz free energy of the system at temperature $T$ and number density $\rho$ can be written as

\begin{equation}
\beta F(T, \rho) = \beta F_{\rm ref}(T, \rho) + \beta F_{\rm bond}(T, \rho)
\end{equation}
\noindent
where $F_{\rm ref}(T, \rho)$ is the free energy of the reference state 
(the state in which no bonds are present) and $F_{\rm bond}(T, \rho)$ is the free-energy contribution due to the bonding between distinct tetramers. In the case of a one-component system of tetrafunctional particles, 
Wertheim's theory provides an expression for the latter in term of
bond probability  $p_b$, defined as  the fraction of formed bonds over the total number of possible bonds, i.e. the probability that one arm of a tetramer is engaged in a bond. The expression reads

\begin{equation}
\beta F_{\rm bond}(T, \rho) = N \left(\log\left[ (1-p_b(T, \rho))^f\right] + \frac{f}{2} p_b(T, \rho) \right)
\end{equation}
\noindent
where $N$ is the number of particles, $f$ is the particle valence ($4$ for tetramers).

Since the free energy depends solely on $f$ and $p_b$, in Wertheim's framework all systems with the same $f$  (and the same reference free energy) share the same thermodynamic behaviour, provided that $p_b$ is used as a scaling variable~\cite{foffikern}. For example, having the hard-sphere fluid as a reference system results in the critical $p_b$ for patchy particles with $f=4$ yielding the value $p_b^c \approx 0.64$, irrespective of the patch shape and interaction strength.

In the case of DNA nanostars, the reference free energy coincides with the free energy of a system  in which the sticky ends bases are scrambled in such a way that Watson-Crick pairing does not occur and no inter-star bonds can form. Under this condition, the residual interaction between different nanostars is essentially provided by
excluded volume, and hence $T$ independent.   
 
To compute the reference free energy we evaluate the equation of state of a system of tetramers at $T = 40 \; ^\circ C$ by simulating tetramers with sticky ends that cannot bind. We then fit the density dependence of the osmotic pressure with the virial expression, estimating in this way the second virial coefficient $B_2$. We find $B_2 = 2100 \pm 190 \; \si{nm}^6$. The corresponding reference free energy is thus
 
\begin{equation}
\beta F_{\rm ref}(T, \rho) = N \left(\log{\rho} - 1 + B_2 \rho^2\right)
\end{equation}
\noindent
where we assume that $B_2$ does not depend on $T$.

The last missing term is the $\rho$ and $T$ dependence of the bond probility $p_b$, a quantity
controlled by the mass action law.  Luckily, accurate estimates of $p_b(T,\rho)$ for 
arbitrary DNA sequences are available in the literature (and on the web), since they enter in 
all oligocalculator programs. In the present case we compute $p_b(T,\rho)$ for the sticky end sequence CGATCG with NUPACK~\cite{nupack}.

 Once an expression for $p_b(T,\rho)$  has been chosen, it becomes possible to 
  locate the spinodal line as the \textit{locus} of points such that the derivative of the pressure $P = \frac{\rho^2}{N} \frac{\partial \beta F}{\partial \rho} $ is zero and the critical point  with the additional condition that
the third derivative vanishes.  

\section{Results}

\subsection{Single tetramer conformation}
The oxDNA model is tailored at reproducing on a quantitative level the structure and the thermodynamics of both single-stranded and double-stranded DNA molecules. In addition, oxDNA can potential be applied to study DNA origamis and other supramolecular DNA assemblies~\cite{oxDNA_review}. Therefore, since the conformation of single tetramers cannot be easily investigated in experiments, computer simulations can provide a valuable tool to predict and inspect the microscopic structure of these DNA constructs. These calculations can then be used as feedback to experimentalists, to be compared with small angle X-ray or neutron scattering to design better building blocks.

\begin{figure}
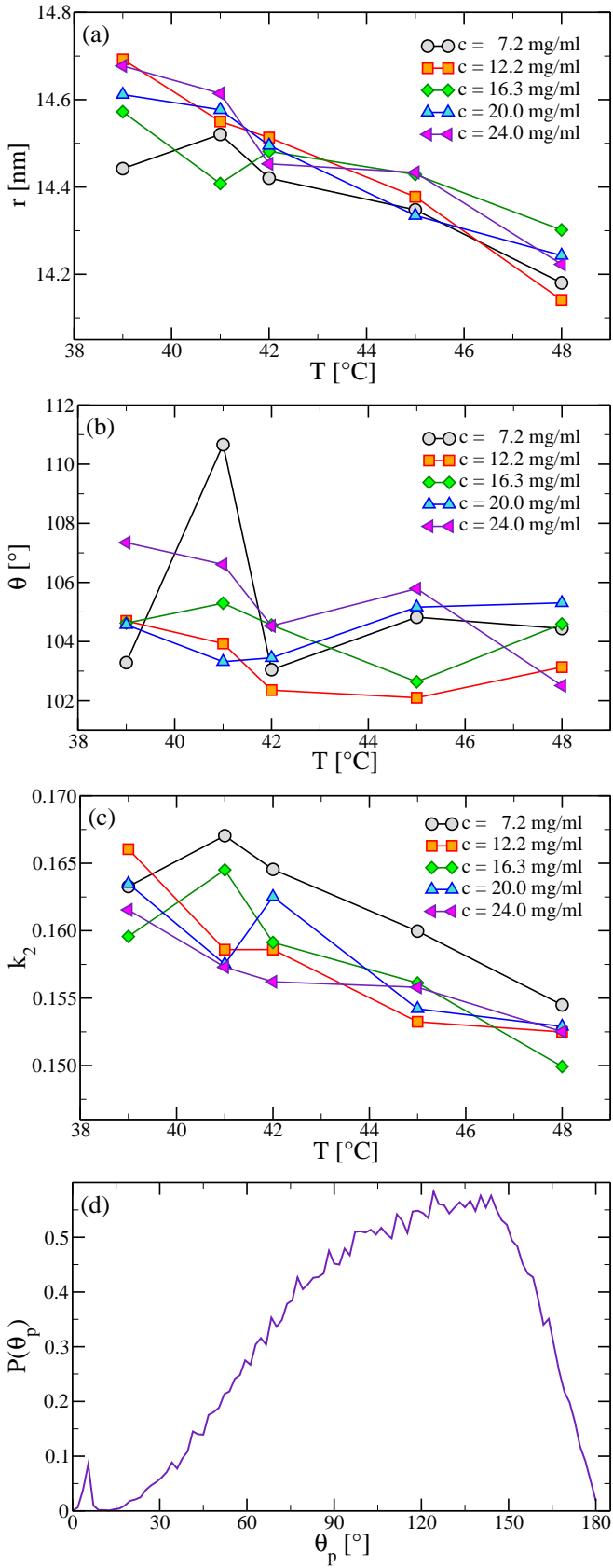

\centering
\includegraphics[width=1\linewidth]{./r.eps}\\
\vspace{0.15cm}
\includegraphics[width=1\linewidth]{./theta.eps}\\
\vspace{0.15cm}
\includegraphics[width=1\linewidth]{./k2.eps}\\
\vspace{0.2cm}
\includegraphics[width=1\linewidth]{./theta_p.eps}\\
\caption{(a) Average distance $r$ and (b) average angle between bonded tetramers $\theta$ for all the investigated state points. The spike in panel (b) is due to statistical noise. (c) Anisotropy parameter as defined in Eq.~\ref{eq:inertia} for all the investigated state points. (d) Probability distribution of the intra-tetramer patch-patch angle $\theta_p$, as defined in the text, for $c=24.0\; \si{mg/ml}$ and $T=39 ^\circ$ C.}
\label{fig:tetra_r_theta}
\end{figure}

We start the analysis by investigating the structure of single tetramers as $T$ and $c$ vary. We look at the distance $r$ between the centres of two bonded tetramers and the angle $\theta$ between triplets of bonded tetramers, averaged over all the configurations at a given $T$ and $c$. The results are presented in Figure~\ref{fig:tetra_r_theta}. For both $r$ and $\theta$ there is no clear trend with $c$ within our numerical error, i.e. the conformation of bonded tetramers does not vary significantly  with concentration in the investigated $c$-range.
Regarding the $T$ dependence, $\theta$ is not affected by $T$, whereas $r$ shows a clear trend (up to a few percent) in the considered $T$-range. This increase in the centre-to-centre distance arises possibly from the stiffening of the central junctions and to the increase of the persistence length as the system is cooled down~\cite{dna_persistence_length}.

To quantify the shape of the nano-star we compute the principal moments of inertia of the tetramers $\lambda_1$, $\lambda_2$ and $\lambda_3$ by diagonalising the tensor of inertia associated to each tetramer. We then define the anisotropy parameter as

\begin{equation}
\label{eq:inertia}
k_2 = 1 - \frac{27 (\lambda_1\lambda_2\lambda_3)}{(\lambda_1 + \lambda_2 + \lambda_3)^3}.
\end{equation}
\noindent
The anisotropy parameter for spherically symmetric ensembles of particles is $k_2=0$, whereas linear chains have $k_2 \to 1$. Planar arrangements, on the other hand, have $k_2 = 5/32 \approx 0.16$~\cite{MayerNature}. $k_2$ for all investigated state points is presented in Fig~\ref{fig:tetra_r_theta}(c). For reference, a tetramer with arms oriented towards the vertices of a perfect tetrahedron yields $k_2\approx 0.012$. Although slightly noisy, the results show that the simulated tetramers tend to be planar rather than on a tetrahedron, a conformation that can also be found in Holliday junctions~\cite{holliday_shape}. As $T$ decreases $k_2$ grows, while changing the concentration does not have a large impact, with only the lowest-$c$ curve having a noticeably higher anisotropy.

We confirm the planar arrangement of the arms by computing the intra-tetramer patch-patch angle $\theta_p$. We first calculate the centres of mass of each tetramer $t$, $\mathbf{r}_t$, and of each of its sticky ends, $\mathbf{r}_t^i$ ($i=1,2,3,4$). We then compute $\theta_p$ as the angle formed by a pair of patches with respect to the tetramer's centre of mass, that is

\begin{equation}
\theta_p(i, j) = \mathrm{acos}\left(\frac{\left(\mathbf{r}_t^i - \mathbf{r}_t\right) \cdot \left(\mathbf{r}_t^j - \mathbf{r}_t\right)}{\left|\mathbf{r}_t^i - \mathbf{r}_t\right| \cdot \left|\mathbf{r}_t^j - \mathbf{r}_t\right|}\right)
\end{equation}
\noindent
Figure~\ref{fig:tetra_r_theta} shows the distribution of the patch-patch angle $P(\theta_p)$ at the highest concentration, $c=24.0\; \si{mg/ml}$, and lowest temperature, $T=39 ^\circ$ C. First of all we note that the distribution is not peaked around the tetrahedral angle $109.5 ^\circ$, but at larger angles, consistent with a non-tetrahedral conformation and with a planar $X$-shaped arrangement of the arms. In addition, the large width of the distribution is another signal of the flexibility of the tetramers. $P(\theta_p)$ also exhibits a peak at very small angles due to bonds between sticky ends belonging to the same tetramer. The number of such intra-tetramer bonds is always very small, of the order of fraction of percent, and decreases with decreasing temperature.

\subsection{Bond probability and average cluster size}

To investigate the static structure of the system and to quantify the extent of the self-assembling process we compute the bond probability $p_b$, the fraction of formed bonds compared to the maximum number of possible bonds (two times the number of tetramers).

\begin{figure}
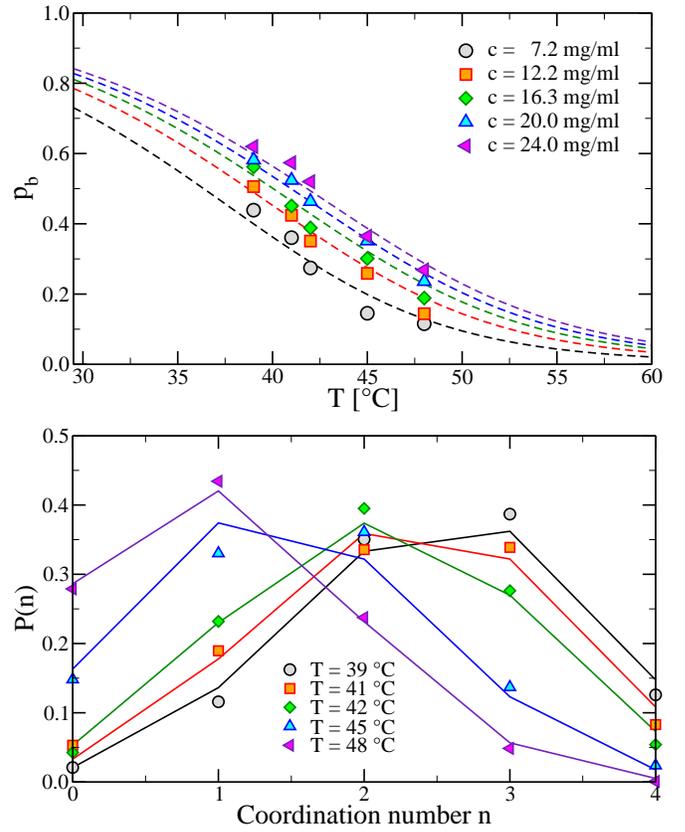

\centering
\includegraphics[width=1\linewidth]{./pb.eps}\\
\vspace{0.15cm}
\includegraphics[width=1\linewidth]{./coordination_number.eps}
\caption{(a) Bond probability $p_b$ as a function of $T$ for all the investigated state points (symbols). The dashed lines are the theoretically estimated melting curve of hexamers of sequence \texttt{CGATCG} (the same as tetramers' sticky ends). (b) Probability for a tetramer to have a coordination number $n$, i.e. $n$ arms bonded, for $c = 24.0 \;\si{mg/ml}$ at all investigated temperatures (symbols) and as predicted by the random percolation theory, Eq.~\ref{eq:perc_theory} (lines).}
\label{fig:tetra_pb}
\end{figure}

Figure~\ref{fig:tetra_pb} shows $p_b$ as a function of $T$ for all the investigated state points.  $p_b$ is a monotonically increasing function of inverse temperature and concentration. We also compare these numerical values with the theoretical melting curves of hexamers having the same sequence as the sticky ends (CGATCG). The theoretical curves, shown as dashed lines in Figure~\ref{fig:tetra_pb}(a), have been computed by NUPACK, a software suite for the analysis and design of nucleic acid systems~\cite{nupack}, using nearest-neighbour empirical parameters~\cite{santalucia} and then calculating free energies and equilibrium concentrations of the DNA in either its single-stranded or double-stranded form. The two  sets of data are rather similar,
suggesting that, in the explored $T$ range, bonds behave in a rather independent way. It also confirms that
this analytic expression for the mass action, consistent with simulation data, is available in the literature
and can be safely compared with numerical and experimental results. We can also use $p_b$ in the framework of percolation theory~\cite{Stauffer} to investigate the connectivity of the system. Figure~\ref{fig:tetra_pb}(b) shows the probability $P(n)$ for a tetramer to have a coordination number $n$, that is, $n$ bonded neighbours, for the high-concentration case. As $T$ decreases the average coordination number increases and non-bonded tetramers become more and more rare. Figure~\ref{fig:tetra_pb}(b) also shows $P(n)$ as predicted by the random bond percolation for four-coordinated particles by the probability distribution function

\begin{equation}
\label{eq:perc_theory}
P(n) = \frac{4!}{n!(4 - n)!} p_b^n(1-p_b)^{4-n}.
\end{equation}
\noindent
The good agreement between numerical and theoretical results suggests that we can safely interpret the ongoing bonding process as a random percolation mechanism.

\begin{figure}
\centering
\includegraphics[width=1\linewidth]{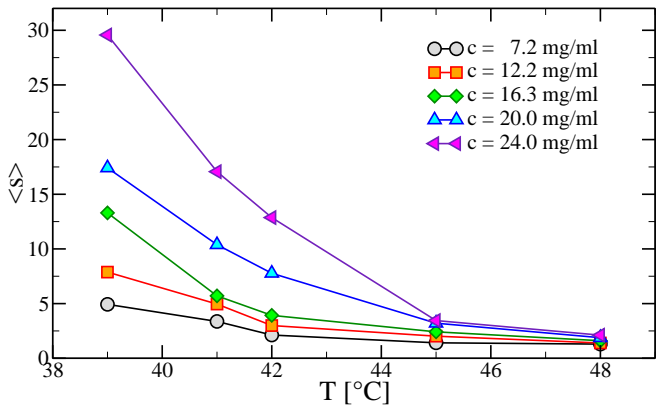}
\caption{Average cluster size $\langle s \rangle$ as a function of $T$ for all the investigated state points.}
\label{fig:tetra_cs}
\end{figure}

As $T$ is lowered, tetramers at all concentrations start to form larger and larger clusters. We investigate the extent of this aggregation by computing the average cluster size $\langle s \rangle$, defined as 
the number of tetramers $N$ divided by the number of clusters $N_c$.  Finite size effects are expected when
$\langle s \rangle$  becomes of the same order as $N$. 

Figure~\ref{fig:tetra_cs} presents $\langle s \rangle$ for all the state points. Similarly to $p_b$, $\langle s \rangle$ increases monotonically with decreasing $T$ or increasing $c$. At the lowest concentration tetramers form small clusters only, and the average cluster size does not change much upon lowering $T$.
By contrast, as $c$ increases $\langle s \rangle$ starts to grow more rapidly with decreasing $T$ and, at the highest concentration $c=24\;\si{mg/ml}$, the average cluster size increases from $\approx 2$ to $\approx 30$ in the investigated region of $T$.

\subsection{Percolation}

\begin{figure}
\centering
\includegraphics[width=1\linewidth]{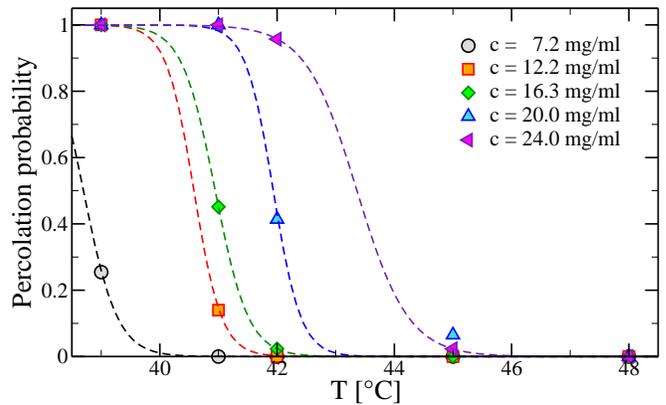}
\caption{Probability percolation for all the investigated concentrations as a function of temperature. The $c = 7.2 \; \si{mg/ml}$, $T = 42 ^\circ C$ value is due to a percolating chain and therefore it is a finite-size effect. Lines are guides for the eye, computed by fitting the numerical data to sigmoidal functions.}
\label{fig:tetra_perc}
\end{figure}

The growth of the cluster brings in a percolation transition, when the cluster size 
becomes comparable to the system size.   In mean-field,
percolation of tetrahedral particles is expected to take place when $p_b =1/3$~\cite{Stauffer}. 
In order to qualitatively estimate the percolation line we start by computing the percolation probability $p_p$, which is defined as the fraction of configurations containing a spanning (infinite) cluster. We detect percolation by veryfing that in the bulk system obtained by
replicating the simulated finite-size box, an infinite size cluster is present. 
Fig.~\ref{fig:tetra_perc} shows $p_p$ against $T$ for all the concentrations.

The percolation probability in finite systems has a sigmoid shape that gets steeper as the system size is increased, eventually becoming a non-analytic  function in the thermodynamic limit. Since the computational cost of a finite-size scaling study is prohibitive, we rely on the results obtained at a single box size. We estimate the percolation temperature $T_{\rm perc}$ by making cuts at $p_p = 0.5$ in Fig.~\ref{fig:tetra_perc}. 
The results are shown in Fig.~\ref{fig:tetra_pd}. We find the bond probability at percolation to be $p_b^p \approx 0.46$ at all concentrations, a value significantly larger than the mean field value $1/3$.   
The larger number of bonds necessary to percolate can be explained by the fact that the mean field value
does not account for the possibility of forming loops of bonds.  In addition, we have noticed that
a small fraction of bonds are involved in double bonds (i.e. pairs of tetramers  with more than one arm bonded),  effectively decreasing the propagation of connectivity. For reference, the previous study of DNA dendrimers predicted $p_b^p$ ranging from $\sim 0.32$ to $p_b \sim 0.42$, depending on concentration~\cite{langmuir}.

\subsection{Structure factor}
\label{subsec:sq}

\begin{figure}
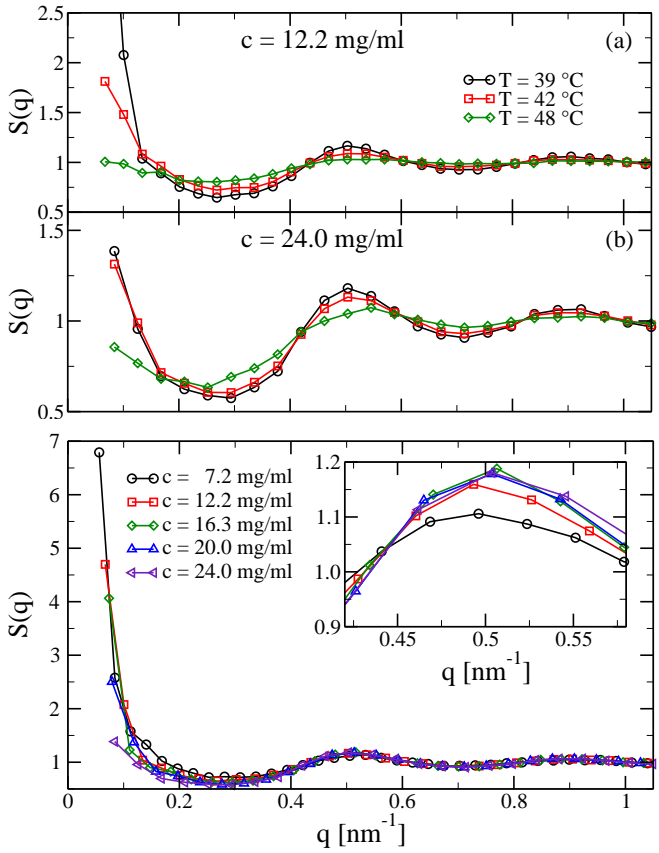

\centering
\includegraphics[width=1\linewidth]{./Sq_c.eps}\\
\vspace{0.12cm}
\includegraphics[width=1\linewidth]{./Sq_T39.eps}	
\caption{Top: static structure factor $S(q)$ for different temperatures for (a) $c = 12.2\; \si{mg/ml}$ and (b) $24.0\; \si{mg/ml}$. The lowest-$q$ value for the $T = 39\; ^\circ C$, $c = 12.2 \; \si{mg/ml}$ $S(q)$ is $\approx 5$ (not shown). Bottom: $S(q)$ at $T = 39 \; ^\circ C$ for all the studied concentrations. The inset shows the first peak in greater detail.}
\label{fig:tetra_sq}
\end{figure}

To provide an estimate of the location of the phase-separation region we study how the static structure factor $S(q)$ evolves with concentration and $T$. The results are shown in Fig.~\ref{fig:tetra_sq}.

Upon lowering $T$, the $S(q)$ becomes more structured, with the height of the peaks and the depth of the minima increasing. The nearest-neighbour peak of the $S(q)$, related to the inter-tetramer bonding, is always located around $q \approx 0.5$, but its position moves slightly towards larger $q$ as the concentration is increased. This value of $q$ corresponds to a real-space distance of $r \approx  12.6\; \si{\nano\metre}$. For comparison, the value obtained by considering two bonded tetramers in the minimum of the energy is $r \approx 14.8 \; \si{\nano\metre}$. As we have seen in Figure~\ref{fig:tetra_r_theta}(a), $r$ approaches this limiting
value on cooling.  
We also note that there is no evidence of a pre-peak, the typical signature of tetrahedral networks.
It has been recently shown that there is a clear correlation between the strength (or the presence) of
the pre-peak and the variance of the center-center-center angle $\theta$ distribution~\cite{ivan_frank_tetra}. For very wide
 $\theta$ distributions (as in the present case) the pre-peak is expected to be missing, since there is no
 sufficient geometrical correlation to establish a tetrahedral network. 

The growth of the low-$q$ limit of  $S(q)$, related to the isothermal compressibility of the system via the relation $S(q\to 0) \propto \chi^{-1}$, signals the development of  structural  heterogeneities.
A divergence of $S(q\to 0)$ is a sign of the approach of a thermodynamic instability (the gas-liquid transition
in our case).   In these cases, the low-$q$ limit of the structure factor provides 
a method to estimate  the critical $T$ or the spinodal temperature $T_s$. Indeed, near a gas-liquid 
spinodal the compressibility factor $\chi$ diverges as a power law with exponent $\gamma = 1.25$, and therefore

\begin{equation}
\label{eq:tetra_extrapolation}
S(q \to 0)^{-1} \propto (T - T_s)^{-\gamma}.
\end{equation}
\noindent

\begin{figure}
\centering
\includegraphics[width=1\linewidth]{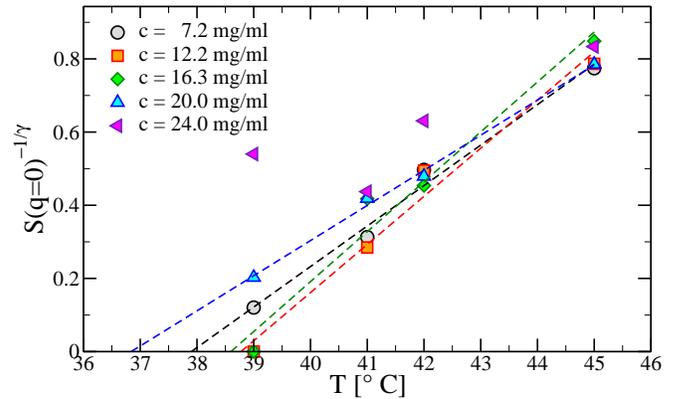}
\caption{Extrapolated scattered intensity to the power of $\gamma$ as a function of $T$ for all the studied concentrations (points). Linearly extrapolating to $0$ (dashed lines) gives the temperatures at which the spinodal is encountered for each value of $c$.}
\label{fig:tetra_extrapolation}
\end{figure}

Experiments performed on the critical isochore showed that the scattered intensity, proportional to
$S(q)$ is well represented by  a Lorentzian  function

\begin{equation}
\label{eq:sql}
S(q)= \frac{S(0)}{1+q^2 \xi ^2}
\end{equation}

where $\xi$ provides a measure of the thermal correlation lenght.  In analogy to the analysis performed
on the experimental data  we fit the low-$q$ part of the structure factor
according to Eq.~\ref{eq:sql}.   From the $T$-dependence of $S(0)$, we find  
 the temperature at which the compressibility diverges according to Eq.~(\ref{eq:tetra_extrapolation}). Unfortunately, the small number of tetramers investigated and the corresponding small simulation box size
 limit the number of small-angle wave-vectors which can be used in the fit.   Fig.~\ref{fig:tetra_extrapolation} shows the numerical results and the obtained fitting curves for all the investigated concentrations but the highest one. The resulting spinodal estimate is reported  in Fig.~\ref{fig:tetra_pd}.

At the largest density investigated the structure factor does not show a divergence on cooling.  
The small angle structure factor grows on cooling but then it saturates to a constant, consistent with the
expectation of an equilibrium gel~\cite{zaccajpcm}.  Indeed, when the system is well beyond the percolation threshold, all particles are part of the same infinite cluster and most of the bonds are formed. Further cooling does not alter the structure of the system and, as a consequence, the topology of the network does not evolve any more~\cite{John,rovigatti_molphys}. 

\subsection{Mean-square displacement}

\begin{figure}
\centering
\includegraphics[width=1\linewidth]{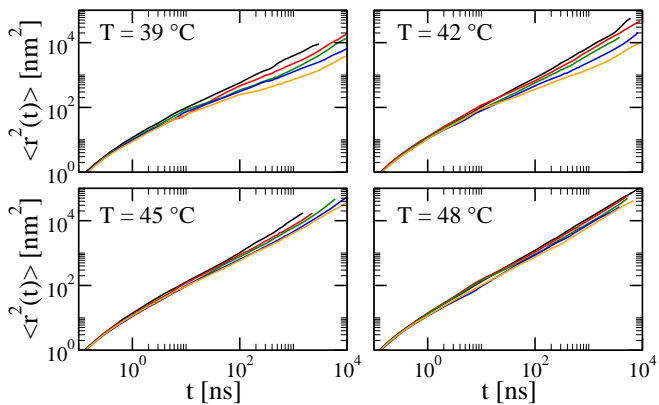}
\caption{Mean-square displacement at $c=7.2 \; \si{mg/ml}$ (black lines), $c=12.2 \; \si{mg/ml}$ (red lines), $c=16.3 \; \si{mg/ml}$ (green lines), $c=20.0 \; \si{mg/ml}$ (blue lines) and $c=24.0 \; \si{mg/ml}$ (orange lines) for four different temperatures (from top left, $T = 39\; ^\circ C$, $42\; ^\circ C$, $45\; ^\circ C$ and $48\; ^\circ C$.}
\label{fig:tetra_msd}
\end{figure}

Next we investigate the dynamics of the system by evaluating the mobility of the tetramers. Fig.~\ref{fig:tetra_msd} shows the mean-square displacement (MSD) $\langle r^2(t) \rangle$ for several concentrations and temperatures. 
At long time the system enters a diffusive regime, where $\langle r^2(t) \rangle \propto t$ and the dynamics 
can be quantified by the diffusion coefficient $D$, defined as

\begin{figure}
\centering
\includegraphics[width=1\linewidth]{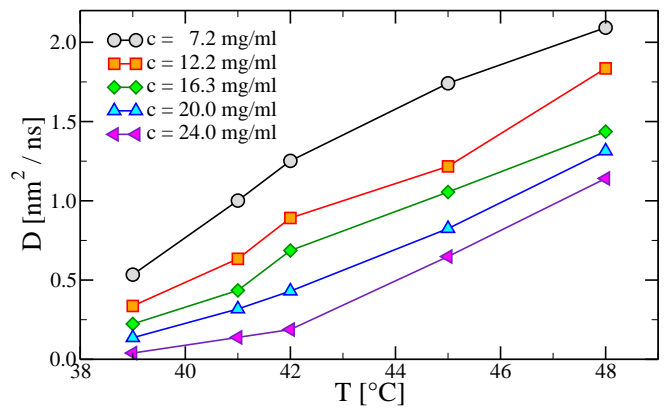}
\caption{Diffusion coefficients for all the investigated state points. Lines are guides for the eye.}
\label{fig:tetra_D}
\end{figure}

\begin{equation}
D = \lim_{t\to \infty} \frac{\langle r^2(t) \rangle}{6t}.
\end{equation}
\noindent
Fig.~\ref{fig:tetra_D} shows $D$ for all the investigated state points.   To access the origin of the
dynamic slowing down, it is useful to draw in the phase diagrams lines of iso-diffusivity, i.e. lines
where the characteristic diffusive time of the particles is comparable. Two of these loci 
are shown in Fig.~\ref{fig:tetra_pd}. 
These lines appear to be parallel to the percolation line (i.e. close to iso-$p_n$ lines), as found in other gel-forming systems~\cite{langmuir,zaccajpcm}. This suggests that the main origin of the slowing down of the dynamics
is the  progressive bond formation.  Finally, we note that at low $T$ the diffusive regime is preceded by  an intermediate sub-diffusive regime, which is more and more pronounced as concentration is increased. Such a sub-diffusive regime possibly signals the onset of the slow dynamics associated with the gel formation. 

\subsection{Phase diagram}

\begin{figure}
\centering
\includegraphics[width=1\linewidth]{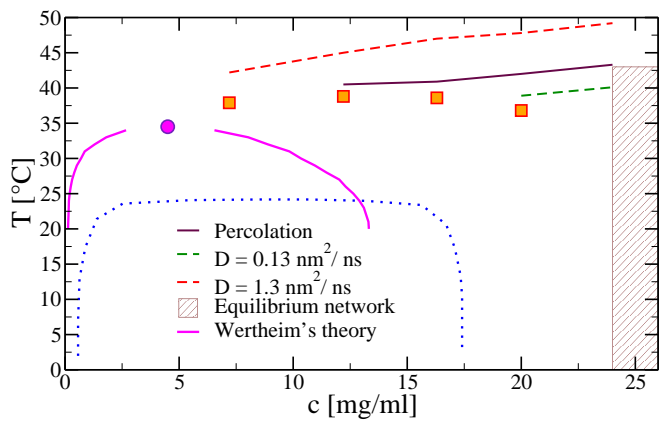}
\caption{Phase diagram of the system. By extrapolating the scattered intensity of all the investigated state points we are able to estimate the location of the spinodal curve (red squares). The blue dotted line is the experimental phase diagram measured at a ionic strength of $50\; \si{mM}$ NaCl~\cite{biffi_dna}. Dashed lines are isodiffusivity curves, while the black solid line is the percolation \textit{locus}. The magenta circle signals the position of the theoretical critical point, while the theoretical spinodal curves are represented as magenta solid lines.}
\label{fig:tetra_pd}
\end{figure}

Fig.~\ref{fig:tetra_pd} shows all the investigated state points and the obtained results, the theoretical phase diagram as computed by Wertheim's theory as well as the experimental phase diagram recently measured experimentally~\cite{biffi_dna}. The experiments have been carried out at the very low salt concentration $50\; \si{mM}$ NaCl, which explains the difference in temperature between the two phase diagrams~\cite{santalucia_review,lifson_salt_DNA}. We note that performing experiments at high salt concentration may be more convenient as the phase separated system is stable well above room temperature.

The difference in concentration, roughly $30\%$, can be similarly explained by noting that it converts to a difference in the inter-tetramer bonding distance  $\approx 10\%$. This change can be qualitatively understood in the framework of Wertheim's theory. Indeed the lower ionic strength enhances the repulsion between the negatively charged DNA strands. This, in turn, effectively increases the contribution to the pressure due to the reference free energy, $\beta F_{\rm ref}$, moving the critical point and the spinodal line to lower concentrations. Even though it has been proven possible to theoretically estimate the salt-dependence of the first virial coefficients of DNA double strands~\cite{stigter1977interactions,nicolai1989ionic,stigter1993theory,hsieh2008ionic}, there are no theoretical approaches to do so for more complicated structures such as these DNA constructs. We plan to experimentally investigate the effect of salt concentration on the phase diagram as a next step towards a deeper understanding of the structure and dynamics of DNA constructs.

We see that Wertheim's theory is able to qualitatively capture the thermodynamic of the system. The theoretical critical temperature, $T_c^w = 34.5 ^\circ$ C, is very close to the numerical one, $T_c \approx 38 ^\circ$ C, with a relative difference in absolute temperature of about $1\%$. On the other hand, the theoretical critical concentration as computed by the Wertheim's theory is roughly half of the numerical one, as observed in similar limited-valence systems~\cite{bian,lungo}. As previously done, we ascribe this difference, which decreases as the valence goes down, to the presence of doubly bonded tetramers and of loops in finite clusters, both of which are not taken into account by Wertheim's theory. The theoretical critical bonding probability comes out to be $p_b^c = 0.52$. In order to estimate the numerical $p_b^c$ we take the concentration having the higher spinodal temperature ($c = 12.2 \; \si{mg/ml}$) as the critical concentration. By doing so we obtain $p_b^c \approx 0.51$, a value very close to the theoretical one.

\section{Conclusions}

In this article we have investigated the phase behaviour and the dynamics of DNA tetramers, i.e. DNA constructs of valence $f=4$. The primary constituents are ssDNA molecules which have been designed to first self-assemble into tetramers and then, upon lowering the temperature, to reversibly form networks of inter-tetramer bonds. This experimental realisation of limited-valence particles is expected to undergo at low temperatures a phase separation between a gas-like diluted phase and a low-concentration, percolating, liquid-like phase. This opens up a region of intermediate concentrations in which the system can be cooled down without encountering any thermodynamic instability. By changing the valence, i.e. by employing constructs with a smaller number of arms, the properties of the resulting thermoreversible gel, an empty liquid~\cite{bian,barbaranatmat}, can be opportunely tuned. 

The present study, featuring very lengthy, large-scale numerical simulations on GPUs, is a step forward in the direction of quantitatively predicting the whole phase diagram of DNA constructs. Indeed, being able to calculate the thermodynamic and dynamic behaviour of complex systems is of paramount importance for designing the materials of tomorrow. By using a sophisticated, realistic model of DNA, simulated on GPUs in order to obtain the required performance speed-up, we have managed, for the first time, to study a bulk system composed of $19600$ nucleotides. We have observed the formation of clusters and, eventually, the appearance of a inhomogeneous percolating network signalling the approaching of the phase separation. We note that the resulting network structure has a tetrahedral nature, as demonstrated by the average value of the angle between bonded triplets of tetramers, even though the tetramers themselves lack a tetrahedral symmetry.

The calculated phase diagram bears similarities with very recent experimental results~\cite{biffi_dna}. Unfortunately, the experiments were performed at a very different salt concentration, making it impossible to quantitatively compare the two phase diagrams. Nevertheless, preliminary results show that carrying out experiments at a salt concentration of $0.5 \UOM{M}$ NaCl increases the spinodal temperature by $10-15 \; ^\circ C$, values which are in line with numerical results. We have also shown that it is possible to compute also some characteristic \textit{locii} like percolation lines and isodiffusivity curves.

We have shown that it is  possible to compute the thermodynamics and dynamic properties of DNA constructs, providing a way of helping in designing novel materials with tunable properties. The reported simulations partly suffer from size limits but show a lot of potential for future applications. Indeed, with the computational power ever increasing, it will be soon possible to perform studies like the present one in a very automated and fast way, so that it will be possible to rely on these results to finely adjust the material properties.

\section*{Acknowledgements}
We acknowledge support from ERC-226207-PATCHYCOLLOIDS, MIUR-PRIN and NVIDIA. LR thanks T.~E. Ouldridge, F. Romano and B.~E.~K. Snodin for discussions.

\end{document}